\documentclass[%
 reprint,
 amsmath,amssymb,
 aps,
]{revtex4-2}

\usepackage{graphicx}
\usepackage{dcolumn}
\usepackage{bm}

\begin{document}

\preprint{APS/123-QED}

\title{Optical fingerprints across the strain-driven semi-Dirac transition in Kekulé-O graphene}

\author{Yawar Mohammadi}%
 \email{y.mohammadi@cfu.ac.ir}
\affiliation{%
 Department of Physics Education, Farhangian University, P.O. Box 14665-889, Tehran, Iran
}%

\date{\today}

\begin{abstract}

We show that the strain-driven semi-Dirac transition in Kekulé-O graphene gives rise to a sequence of anisotropic optical fingerprints associated with band structure reconstruction. Across the transition, optical spectral weight is continuously redistributed among the dominant interband transitions, leading to a pronounced enhancement of the optical anisotropy. Combining numerical four-band calculations with analytical low-energy results, we identify three low-energy fingerprints that emerge with increasing strain: gapped absorption peaks, semi-Dirac critical scaling, and a pronounced van Hove optical resonance. At the semi-Dirac critical point, where the Kekulé gap closes at the $\Gamma$ point, the low-energy optical conductivity is characterized by $\sigma_{xx}(\Omega)\propto\Omega^{1/2}$ and $\sigma_{yy}(\Omega)\propto\Omega^{-1/2}$. Beyond the transition, the semi-Dirac point splits into two anisotropic Dirac cones, accompanied by the emergence of saddle points near the $\Gamma$ point. The resulting saddle-point excitations produce a pronounced van Hove optical resonance at energies well below those of graphene, while the split Dirac cones give rise to an anisotropic constant optical conductivity. We further show that the low-energy optical fingerprints can be traced to the continuous evolution of a dominant optical transition channel driven by strain-induced band reconstruction. Moreover, the fingerprints remain identifiable in the presence of moderate disorder broadening and finite-temperature effects, indicating their potential observability under experimentally realistic conditions.

\end{abstract}

\maketitle


\section{\label{sec:Introduction}Introduction}

Recent experiments have shown that lattice reconstructions in graphene can profoundly modify its energy spectrum through intervalley coupling between the inequivalent $\mathbf{K}$ and $\mathbf{K}'$ valleys. In particular, Bao \textit{et al.}~\cite{Bao1} reported an O-shaped Kekulé modulation (Kekulé-O) in lithium-intercalated graphene, establishing a platform in which intervalley hybridization opens a finite gap at the Dirac point and gives rise to massive Dirac fermions. A Y-shaped Kekulé modulation (Kekulé-Y) has also been observed in graphene on copper substrates~\cite{Gutierrez1}. It has been shown theoretically that these two Kekulé phases lead to qualitatively different low-energy electronic structures~\cite{Gamayun1}: Kekulé-O produces a gapped Dirac spectrum through a mass-generating term, whereas Kekulé-Y preserves two species of gapless Dirac fermions with valley-momentum locking.

The electronic properties of Kekulé graphene have been extensively investigated in recent years~\cite{Andrade1,Wang1,Guan1,Santacruz1,Tijerina1}. In particular, the Kekulé-O phase has attracted considerable attention as a platform for massive Dirac fermions and tunable band engineering. Previous studies have explored Floquet band reconstruction~\cite{Mojarro1}, optically driven Josephson effects~\cite{Zeng1}, flat-band formation in moiré and twisted systems~\cite{Scheer1}, and domain-wall confinement~\cite{Garcia2}. In comparison, the Kekulé-Y phase has been mainly studied in the context of transport and valley-contrasting phenomena~\cite{Wu1,Andrade2,Wang2,Iurov1,Garcia1,Li1}, optical responses~\cite{Herrera1,Mohammadi1,SantacruzG1,Mohammadi2}, integer quantum Hall effects~\cite{Mohammadi3}, and collective phenomena~\cite{Herrera2,Alimohammadi1}. Despite this broad activity, the optical response of Kekulé-O graphene under strain has remained largely unexplored.

In low-dimensional materials, strain provides a powerful and reversible route for engineering the electronic structure, driving substantial band reconstruction and the emergence of distinct electronic regimes~\cite{Naumis1,Si1,Pereira1}. As a result, it enables direct control over a wide range of properties, including transport~\cite{CastroNeto1,Mohammadi4}, optical responses~\cite{Ni1,Chhikara1}, collective carrier dynamics~\cite{Pellegrino1}, and magneto-transport phenomena arising from pseudomagnetic fields~\cite{Levy1,Guinea1}.

In Kekulé-modulated graphene, strain plays a nontrivial role by competing with the Kekulé order and modifying the valley-coupled electronic structure. Unlike Kekulé-Y graphene, where strain mainly induces valley-dependent gauge fields~\cite{Andrade3,Mohammadi5}, the Kekulé-O phase exhibits a qualitatively different response, with a continuous strain-driven evolution of the spectrum that, in this work, is shown to culminate in a semi-Dirac transition.

Motivated by this strain-driven band reconstruction, we investigate how the optical response evolves as a function of strain and reveal a sequence of distinct strain-induced regimes. While strain effects and Kekulé ordering have been widely studied separately, a unified description of optical conductivity in Kekulé-O graphene, particularly across the semi-Dirac transition, remains lacking. In this work, we develop a combined numerical and analytical framework to systematically characterize the strain-dependent optical conductivity and identify its associated optical fingerprints.

We show that across increasing strain, the optical response exhibits distinct anisotropic fingerprints. At weak strain, the system exhibits a gapped absorption peak, followed by a semi-Dirac critical point characterized by anisotropic optical conductivity with power-law scaling. At higher strain, the semi-Dirac point splits into two anisotropic Dirac cones and saddle points develop near the $\Gamma$ point. The resulting saddle-point excitations generate a pronounced van Hove optical resonance at energies well below those of graphene, while the split Dirac cones give rise to an anisotropic constant optical conductivity. Overall, the low-energy features form a unified sequence of strain-controlled optical fingerprints arising from energy spectrum reconstruction, which remain identifiable under moderate disorder and finite-temperature effects, indicating their observability under experimentally realistic conditions.

The remainder of this paper is organized as follows. In Sec.~\ref{sec:Model}, we introduce the Hamiltonian for Kekulé-O graphene and discuss the corresponding band structure together with the optical conductivity formalism. Section~\ref{sec:Results} presents the numerical results obtained from the full four-band model and analyzes the evolution of the optical response across the different strain regimes. We further derive analytical expressions for the anisotropic constant optical conductivity across these regimes, providing a detailed characterization of the optical response. Finally, Sec.~\ref{sec:Conclusions} summarizes the main results of the work. Additional technical details, which outline the construction of the low-energy Hamiltonian and the derivation of the analytical optical conductivity, are given in Appendices~\ref{AppendixA} and ~\ref{AppendixB}.

\section{\label{sec:Model}Model and Formalism}

Starting from the continuum Hamiltonian introduced in Ref.~\cite{Andrade3}, we investigate the strain-driven reconstruction of the energy spectrum and its consequences for the optical conductivity in Kekulé-O graphene. Within this framework, the optical conductivity is formulated using the Kubo formalism for the strained system across different electronic regimes.

\begin{figure}
\includegraphics[width=7.5cm,angle=0]{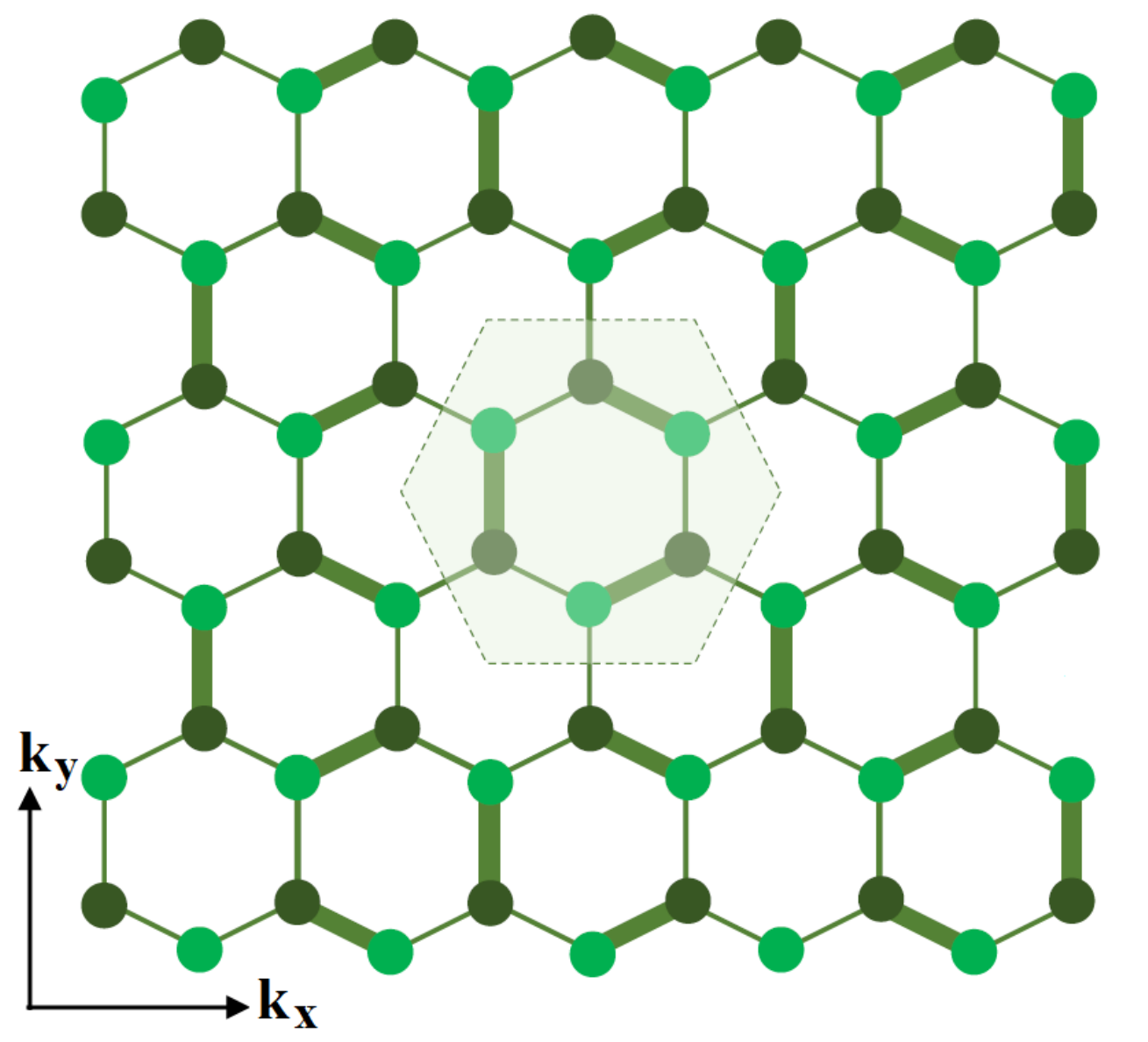}
\caption{\label{Fig01} Schematic of graphene with a uniform Kekulé-O bond modulation. The thick and thin bonds represent alternating strong and weak hopping amplitudes characteristic of the Kekulé-O pattern. The shaded hexagon denotes the enlarged unit cell of the distorted lattice.}
\end{figure}

\vspace{0.5cm}
\textbf{Model Hamiltonian and energy spectrum.} We consider the continuum effective Hamiltonian describing strained Kekulé-O graphene~\cite{Andrade3}:
\begin{equation}\label{Eq01}
 \hat{H}=\left(
             \begin{array}{cc}
   \hbar v_{F}\bm{\hat{\sigma}}.(\mathbf{\bar{k}}+\mathbf{A})   &  \hat{H}_{12}  \\
  \hat{H}_{12}^{\dagger}   &   \hbar v_{F}\bm{\hat{\sigma}}.(\mathbf{\bar{k}}-\mathbf{A})
                \end{array}
                       \right).
\end{equation}
where the off-diagonal block responsible for intervalley coupling is given by
\begin{equation}\label{Eq02}
\hat{H}_{12} = \Delta(3\tilde{t}_{0}\hat{\sigma}_{z}-i\hbar v_{F}a(\mathbf{A}\times\mathbf{k})_{z}\hat{\sigma}_{0}).
\end{equation}
Here, $ v_{F}=3 a t_{0}/2\hbar $, with $ t_{0}=3~\mathrm{eV} $ and $ a=1.42~\mathrm{\AA} $ the nearest-neighbor hopping energy and carbon-carbon bond length of pristine graphene, respectively. The Pauli matrices $\hat{\sigma}_{i}$ ($i \in \{0,x,y,z\}$) act in sublattice space, $ \Delta $ is the Kekul\'{e} coupling amplitude, and \(\mathbf{k}\) denotes the crystal momentum.

Compared to the Hamiltonian of Kekulé-O graphene in the absence of strain~\cite{Gamayun1}, the application of uniaxial strain results in three primary modifications. First, in the diagonal block of the Hamiltonian, strain modifies the crystal momentum~\cite{Oliva1}, leading to $\mathbf{\bar{k}} = (\hat{1} + (1 - \beta)\mathbf{\varepsilon})\mathbf{k}$, which governs the anisotropic deformation of the Dirac dispersion. Here, $\beta\approx3$ is the Gr\"uneisen parameter accounting for strain-induced corrections to the hopping amplitudes~\cite{Pereira1}, $\hat{1}$ denotes the $2\times2$ identity matrix, and $\hat{\varepsilon}$ is the strain tensor. For a uniaxial deformation of magnitude $\varepsilon$ applied at an angle $\theta$ with respect to the zigzag (x) direction, the strain tensor is given by
\begin{equation}\label{Eq03}
  \hat{\varepsilon}=\left(
             \begin{array}{cc}
  \varepsilon(\cos^{2}\theta-\nu \sin^{2}\theta)   &      \varepsilon(1+\nu)\cos\theta\sin\theta      \\
     \varepsilon(1+\nu)\cos\theta\sin\theta        &   \varepsilon(\sin^{2}\theta-\nu \cos^{2}\theta)
                \end{array}
                       \right),
\end{equation}
with Poisson ratio $\nu\approx0.165$~\cite{Blakslee1}. Second, strain induces a pseudovector potential~\cite{Andrade3}, with components expressed in terms of the strain tensor as $\mathbf{A}=\left(\frac{\beta}{2a}(\varepsilon_{xx}-\varepsilon_{yy}), -\frac{\beta}{2a}\varepsilon_{xy}\right)$, which leads to a modification of the Dirac valleys in momentum space. Third, it modifies the nearest-neighbor hopping amplitude according to $\tilde{t}_{0}=t_{0}[1-\frac{\beta\varepsilon}{2}(1-\nu)]$, which contributes to the strain-induced renormalization of the gap and band structure.

Diagonalizing Eq.~(\ref{Eq01}) yields the energy band structure
\begin{widetext}
\begin{eqnarray}\label{Eq04}
E^{\lambda,s}_{\mathbf{k}} &=& \lambda \Bigg(~\hbar^{2} v_{F}^{2}(|\bm{\bar{k}}|^{2}+|\mathbf{A}|^{2})+\Delta^{2}
\Big([3\tilde{t}_{0}]^{2}+[\hbar v_{F}a(\mathbf{A}\times\mathbf{k})_{z}]^{2} \Big) + 2s\hbar v_{F} \Bigg[ \hbar^{2}
v_{F}^{2}(\mathbf{A}.\mathbf{\bar{k}})^{2} \nonumber \\
 &+& \Delta^{2}\Big(~\hbar^{2}v_{F}^{2}a^{2}|\mathbf{\bar{k}}|^{2}(\mathbf{A}\times \mathbf{k})_{z}^{2} + 2\hbar v_{F}a
 (3\tilde{t}_{0})(\mathbf{A}\times \mathbf{k})_{z}(\mathbf{A}\times \mathbf{\bar{k}})_{z} + (3\tilde{t}_{0})^{2}|
 \mathbf{A}|^{2}~\Big) \Bigg]^{1/2} ~  \Bigg)^{1/2},
\end{eqnarray}
\end{widetext}
where $\lambda=\pm1$ denotes conduction and valence bands, and $s=\pm1$ labels the strain-split subbands.

\begin{figure*}
\centering
\includegraphics[width=16cm,angle=0]{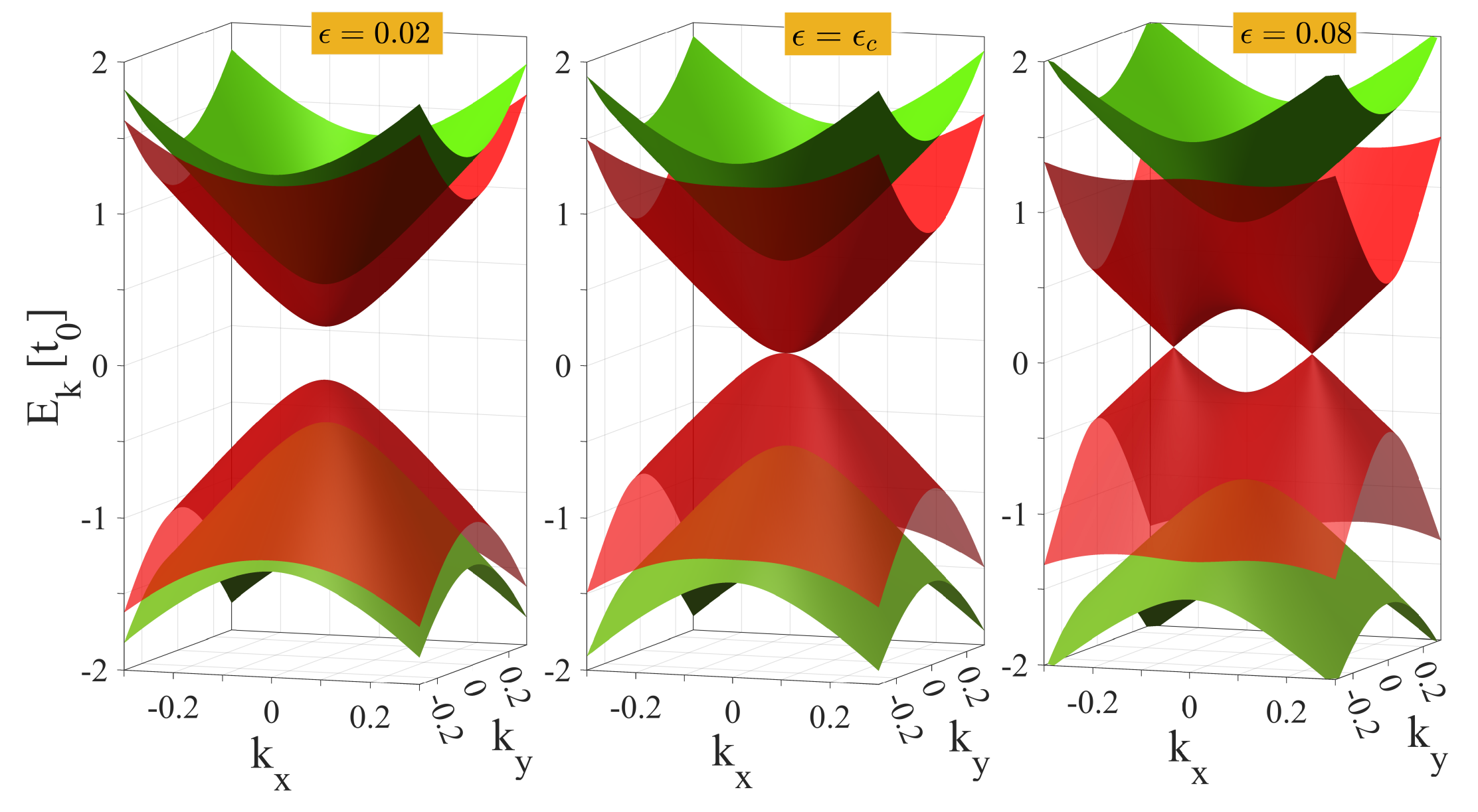}
\caption{\label{Fig02} Evolution of the band structure of Kekulé-O graphene with fixed Kekulé modulation strength $\Delta=0.04$ under uniaxial strain applied along the $x$ direction for $\varepsilon=0.02$, $\varepsilon=\varepsilon_c$, and $\varepsilon=0.08$. Strain lifts the degeneracy of the unstrained bands, driving one pair toward zero energy and the other to higher energies. At $\varepsilon=\varepsilon_c$, the gap closes and a semi-Dirac spectrum emerges. For $\varepsilon>\varepsilon_c$, the semi-Dirac node splits into two anisotropic Dirac cones, accompanied by the formation of saddle points near the $\Gamma$ point, while the higher-energy bands move further away from the low-energy sector.}
\end{figure*}

Figure~\ref{Fig02} shows the evolution of the band structure of strained Kekulé-O graphene for three representative values of uniaxial strain applied along the zigzag ($x$) direction. In the \textit{gapped regime} ($\varepsilon = 0.02$), the band structure is modified compared to the unstrained case, accompanied by a lifting of band degeneracies. As the strain increases, the band gap at the $\Gamma$ point progressively decreases and closes at the critical strain $\varepsilon_c$, where a \textit{semi-Dirac critical point} emerges with a gapless spectrum. For $\varepsilon > \varepsilon_c$, the system enters a \textit{post-critical regime} in which the semi-Dirac node splits into two anisotropic Dirac cones located at $\pm K_{D}$ along the $k_x$ direction, reconstructing the band topology. At the same time, the band extrema near the $\Gamma$ point evolve into saddle points, producing a van Hove singularity in the electronic spectrum.

\begin{figure*}
\centering
\includegraphics[width=16cm,angle=0]{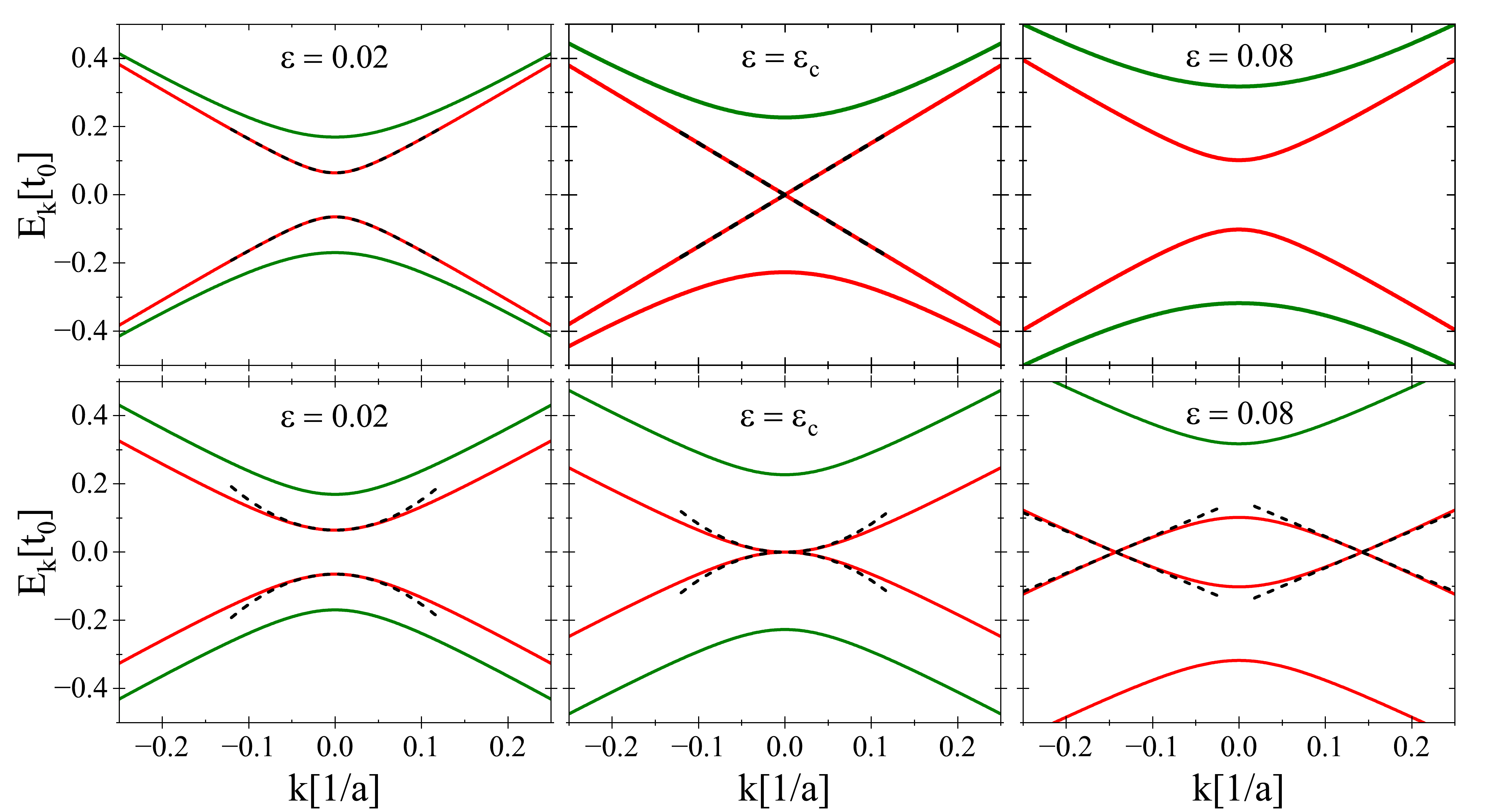}
\caption{\label{Fig03} Band-structure cuts along $k_x = 0$ (top panels) and $k_y = 0$ (bottom panels) for strain values $\varepsilon = 0.02$, $\varepsilon_c$, and $0.08$, corresponding to uniaxial deformation along the zigzag direction. The horizontal axis denotes $k_y$ (top panels) and $k_x$ (bottom panels), respectively. Solid lines represent the full continuum model, while dashed lines show the results obtained from the effective low-energy Hamiltonian (Appendix~\ref{AppendixA}). Good agreement between the two descriptions is observed in the low-energy regime, particularly near the $\Gamma$ point and around the new Dirac points in the post-critical regime.}
\end{figure*}

To further illustrate the anisotropic character of the dispersion, Fig.~\ref{Fig03} shows band-structure cuts along $k_x = 0$ and $k_y = 0$ for the same strain values. These cross sections highlight the pronounced anisotropy of the strain-induced spectrum. Solid lines denote the full continuum model, whereas dashed lines correspond to the effective low-energy Hamiltonian (Appendix~\ref{AppendixA}). The close agreement between the two descriptions demonstrates the validity of the low-energy model over the relevant energy range, including the vicinity of the $\Gamma$ point in the gapped regime and at the semi-Dirac critical point, as well as the post-critical regime characterized by anisotropic Dirac cones.

\vspace{0.5cm}
\textbf{Optical conductivity.}
The optical conductivity of a system can be computed using the Kubo formula. In general, the real (absorptive) part of the longitudinal conductivity at finite temperature, $T$, and photon energy, $\hbar\Omega$, can be written as~\cite{Nicol1,Carbotte1}
\begin{eqnarray}\label{Eq05}
\sigma_{\alpha\alpha}(\Omega,T) &=& \frac{g_s \hbar e^2}{2\Omega}
\int_{-\infty}^{+\infty} \frac{d\omega}{2\pi} [n(\omega) - n(\omega + \Omega)] \nonumber \\
& & \int \frac{d^2 k}{4\pi^2} \mathrm{Tr} \big[ \hat{v}_\alpha \, \mathcal{\hat{A}}(\mathbf{k}, \omega)
\hat{v}_\alpha  \mathcal{\hat{A}}(\mathbf{k}, \omega + \Omega) \big],
\end{eqnarray}
where $g_s$ is the spin degeneracy, $e$ is the electron charge, and $ n(\omega) = [ e^{\beta (\hbar\omega - \mu)} + 1 ]^{-1} $ is the Fermi-Dirac distribution function, with $ \beta = \frac{1}{k_B T} $, and $\mu$ the chemical potential measured from the charge neutrality point. The velocity matrix is given by $\hat{v}_\alpha = \frac{1}{\hbar}\frac{\partial \hat{H}}{\partial k_\alpha} $ and the spectral function matrix is obtained from the retarded Green's function matrix as $\mathcal{\hat{A}}(\mathbf{k}, \omega) = -2 \, \mathrm{Im} \hat{G}(\mathbf{k}, \omega + i 0^+)$.

In order to evaluate the contribution of individual energy bands to the optical conductivity, the spectral function matrix is expressed in terms of projection operators $\hat{M}^{\lambda,s}(\mathbf{k})$ associated with each band $(\lambda,s)$ as
\begin{equation}\label{Eq06}
\mathcal{\hat{A}}(\omega, \mathbf{k}) = 2\pi \sum_{\lambda, s}
\hat{M}^{\lambda, s}(\mathbf{k}) \,
\delta(\hbar \omega - E_{\mathbf{k}}^{\lambda, s}).
\end{equation}
Substituting this expression into Eq.~(\ref{Eq05}), the absorptive part of the optical conductivity can be written as~\cite{Mohammadi5}
\begin{eqnarray}\label{Eq07}
\sigma_{\alpha\alpha}(\Omega,T) &=& \frac{\pi g_s e^2}{\Omega}
\sum_{\lambda,\lambda',s,s'} \int \frac{d^2 k}{4\pi^2}
\left[ n_{\mathbf{k}}^{\lambda,s} - n_{\mathbf{k}}^{\lambda',s'} \right] \nonumber \\
& & \delta \left(\hbar \Omega + E_{\mathbf{k}}^{\lambda,s} - E_{\mathbf{k}}^{\lambda',s'} \right)
P_{\alpha\alpha}^{\lambda,s \to \lambda',s'}(\mathbf{k}),
\end{eqnarray}
where the squared velocity matrix elements are given by
\begin{equation}\label{Eq08}
P_{\alpha\alpha}^{\lambda,s \to \lambda',s'}(\mathbf{k}) =
\mathrm{Tr} \left[ \hat{v}_{\alpha} \, \hat{M}^{\lambda', s'}(\mathbf{k}) \,
\hat{v}_{\alpha} \, \hat{M}^{\lambda, s}(\mathbf{k}) \right].
\end{equation}

The optical conductivity is governed by the interplay between the energy-conserving phase space encoded in the Dirac delta function and the squared velocity matrix elements \(P_{\alpha\alpha}^{\lambda,s \to \lambda',s'}(\mathbf{k})\). The former determines the number of states satisfying the optical resonance condition and is closely connected to the joint density of states (JDOS), whose enhancement near extrema or saddle points of the band structure can produce pronounced optical features. The velocity matrix elements, on the other hand, encode the momentum-dependent pseudospin structure of the electronic states and determine the strength of the optical transitions. Depending on the overlap between the initial and final states, specific interband processes may be enhanced or strongly suppressed, leading to characteristic polarization-dependent optical responses. Such matrix-element effects are well known in graphene-based systems and can give rise to effective optical selection rules associated with the underlying pseudospin and chirality of the low-energy states~\cite{Herrera1,CastroNeto1,Chung1,Pereira2}. Together, the JDOS and velocity matrix elements provide the framework for interpreting the strain-dependent optical conductivity discussed in the following section.

\section{\label{sec:Results} Results and discussion}

In this section, we present our results for the longitudinal optical conductivity of Kekulé-O graphene under uniaxial strain across the strain-induced regimes shown in Fig.~\ref{Fig02}, including the gapped regime, the semi-Dirac critical point, and the post-critical regime. For the numerical implementation of the terms containing Dirac delta functions, we use the Lorentzian representation $ \delta(x)=\frac{\Gamma/\pi}{x^{2}+\Gamma^{2}} $, where $\eta$ denotes the phenomenological broadening parameter. Unless otherwise stated, all calculations are performed at $\eta=0.001t_0$, zero temperature, and zero chemical potential, corresponding to the charge-neutrality point.

\begin{figure*}
\centering
\includegraphics[width=15cm,angle=0]{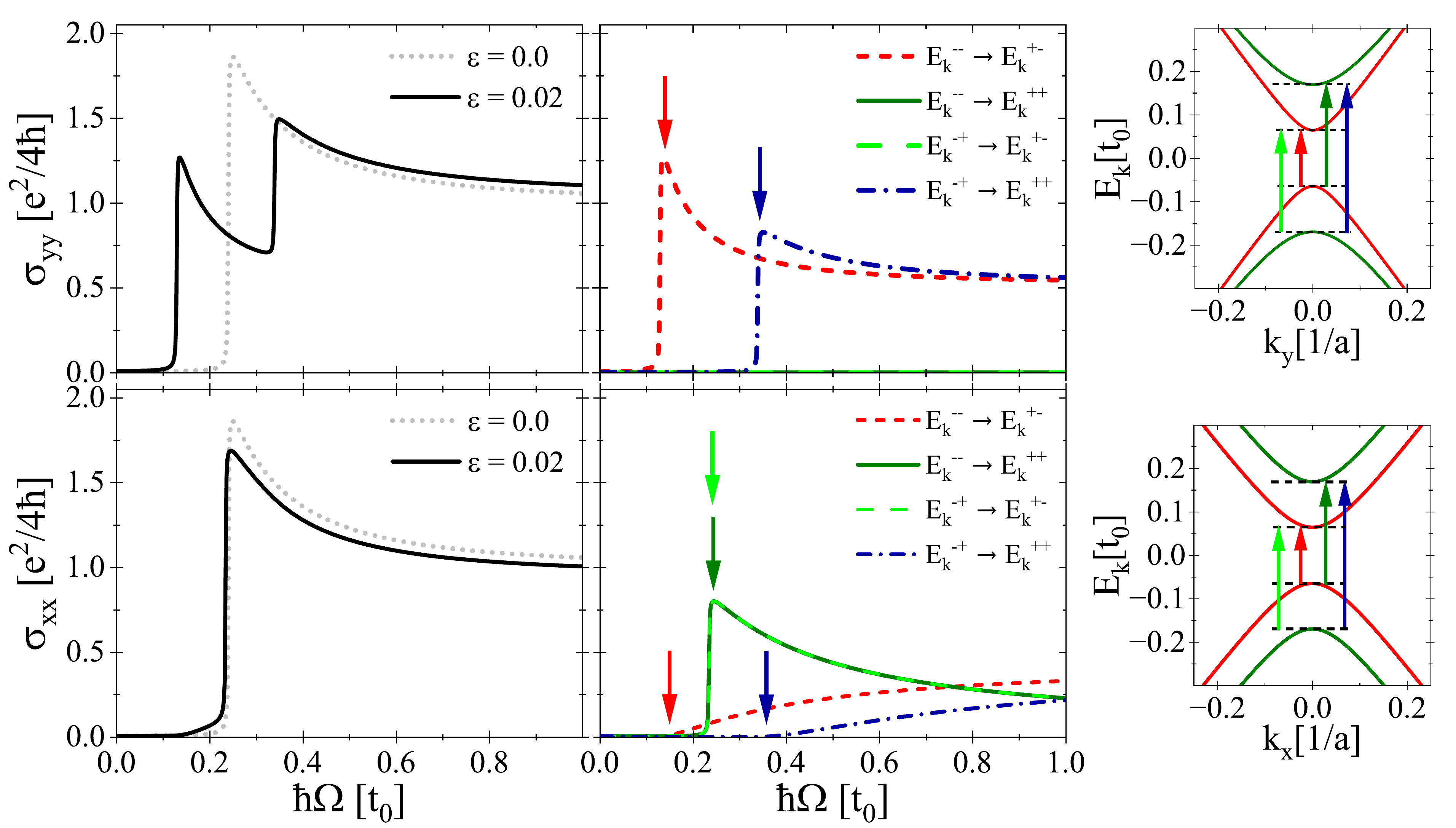}
\caption{\label{Fig04} Longitudinal optical conductivity of strained Kekulé-O graphene in the gapped regime under uniaxial strain $\varepsilon=0.02$ applied along the zigzag ($x$) direction. The left panels show $\sigma_{yy}$ (top) and $\sigma_{xx}$ (bottom), with dashed gray curves indicating the unstrained case. The middle panels resolve the response into contributions from intra-$s$ and inter-$s$ interband transitions, while the right panels display the corresponding band structures along $k_y$ and $k_x$ with arrows indicating the dominant optical transitions.}
\end{figure*}

Figure~\ref{Fig04} illustrates the longitudinal optical conductivity of Kekulé-O graphene in the gapped regime under uniaxial strain $\varepsilon=0.02$ applied along the zigzag ($x$) direction. The left panels show $\sigma_{xx}$ and $\sigma_{yy}$ together with the corresponding unstrained results (gray dotted curves). The middle panels resolve the optical conductivity into intra-$s$ and inter-$s$ transition channels, namely $E_{\mathbf{k}}^{--}\rightarrow E_{\mathbf{k}}^{+-}$ and $E_{\mathbf{k}}^{-+}\rightarrow E_{\mathbf{k}}^{++}$ (intra-$s$), and $E_{\mathbf{k}}^{--}\rightarrow E_{\mathbf{k}}^{++}$ and $E_{\mathbf{k}}^{-+}\rightarrow E_{\mathbf{k}}^{+-}$ (inter-$s$). The right panels display the energy band dispersion along $k_y$ and $k_x$, with arrows indicating the optically allowed interband transitions.

In the absence of strain, the system exhibits an isotropic optical response characterized by an absorption onset at the gap energy and an approximately constant high-energy conductivity, consistent with gapped Dirac systems~\cite{Gusynin1,Gusynin2}. Under uniaxial strain, the optical spectral weight is redistributed among the available interband channels, leading to anisotropic responses in $\sigma_{xx}$ and $\sigma_{yy}$. For $y$ polarization, the response is dominated by intra-$s$ transitions and exhibits a double-peak structure followed by a nearly constant optical conductivity at higher energies. In contrast, for $x$ polarization both intra-$s$ and inter-$s$ channels contribute, with inter-$s$ transitions governing the absorption onset and producing the main low-energy feature in $\sigma_{xx}$. At high energies, all active channels contribute, leading to an nearly constant optical conductivity with an asymptotic value different from that of $\sigma_{yy}$.

This anisotropy originates from the strain-induced reconstruction of the electronic spectrum (right panels). Along the $k_y$ direction, the band structure largely retains its gapped Dirac-like character, leading to strongly suppressed velocity matrix elements for inter-$s$ transitions. As a result, $\sigma_{yy}$ is governed almost entirely by the intra-$s$ channels. The resulting double-peak structure arises from two distinct gapped absorption thresholds associated with these intra-$s$ transition channels, and is further enhanced by the JDOS near the corresponding absorption edges. In contrast, the stronger reconstruction along the $k_x$ direction substantially modifies the pseudospin texture, thereby lifting the suppression of inter-$s$ velocity matrix elements and generating a finite overlap between states belonging to different $s$-labeled bands. Consequently, the inter-$s$ transitions acquire sizeable velocity matrix elements due to strain-induced band mixing and contribute significantly to $\sigma_{xx}$. Further enhanced by the JDOS at the absorption edge, these transitions govern the dominant low-energy absorption feature in $\sigma_{xx}$. Moreover, the approximately constant $\sigma_{xx}$ and $\sigma_{yy}$, arising from the combined contributions of all active channels, reflects the recovery of an anisotropic Dirac-like dispersion away from the gap region~\cite{Mohammadi5}.

\begin{figure*}
\centering
\includegraphics[width=15cm,angle=0]{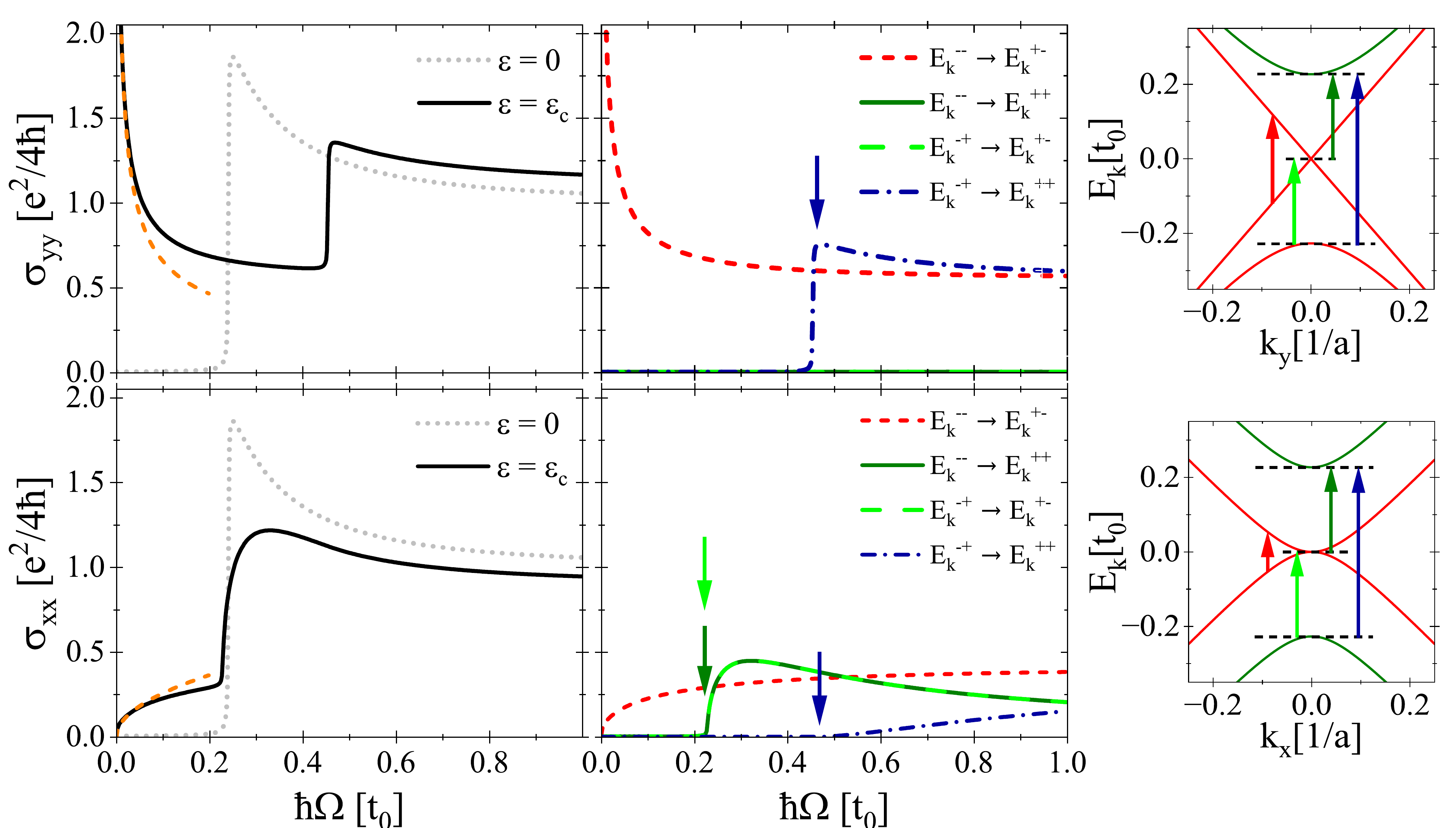}
\caption{\label{Fig05} Optical conductivity of strained Kekulé-O graphene at the critical strain $\varepsilon=\varepsilon_c$, where the system undergoes a transition to a semi-Dirac state with mixed linear and quadratic dispersion. The left panels show $\sigma_{yy}$ (top) and $\sigma_{xx}$ (bottom); dashed gray curves denote the unstrained case, while red dashed-dotted curves represent the analytical low-energy theory derived in Appendix B. The middle panels decompose the response into intra-$s$ and inter-$s$ interband transitions, and the right panels show the corresponding band dispersions along $k_y$ and $k_x$ with transition-resolved arrows. The results reveal a collapse of the finite-frequency absorption gap and a redistribution of spectral weight consistent with semi-Dirac scaling behavior.}
\end{figure*}

Figure~\ref{Fig05} presents the longitudinal optical conductivity at the semi-Dirac critical point, where the Kekulé-induced gap closes and the low-energy spectrum acquires a linear dispersion along the $k_y$ direction and a quadratic dispersion along $k_x$. The middle and right panels show the decomposition of the optical conductivity into intra-$s$ and inter-$s$ transition channels and the reconstructed band structure, respectively, following the notation of Fig.~\ref{Fig04}.

The collapse of the absorption threshold at $\varepsilon=\varepsilon_c$ gives rise to a critical low-energy optical response. As a consequence, in the zero-energy limit the response becomes strongly anisotropic: $\sigma_{yy}$ is markedly enhanced, whereas $\sigma_{xx}$ is suppressed. Moreover, $\sigma_{yy}$ remains dominated by intra-$s$ channels, exhibits a peak associated with the second intra-$s$ transition, and saturates to a constant value at high energies. In contrast, for $x$ polarization, both intra-$s$ and inter-$s$ transitions contribute to $\sigma_{xx}$, with the low-energy continuum governed by the lowest intra-$s$ channel. The higher-energy part of the spectrum shows a weakening of absorption features, while the remaining intra-$s$ channel shifts toward higher excitation energies. At high energies, $\sigma_{xx}$ also approaches a distinct, constant asymptotic value.

The origin of this anisotropic low-energy response can be traced to the emergence of a gapless semi-Dirac spectrum at the $\Gamma$ point. The resulting band structure produces a singular low-energy JDOS together with strongly anisotropic velocity matrix elements. For $\sigma_{yy}$, the linear dispersion along $k_y$ keeps the velocity matrix elements finite [see Eq.~(\ref{EqB06})], so that the singular JDOS directly governs the optical response, leading to the characteristic divergence $\sigma_{yy}\propto\Omega^{-1/2}$. This behavior is captured by the low-energy effective Hamiltonian derived in Appendix~\ref{AppendixA}, which yields the optical response detailed in Appendix~\ref{AppendixB}:
\begin{equation}\label{Eq09}
\sigma_{yy}(\Omega)=
\frac{e^{2}}{4\hbar}
\frac{g_sG}{3}
\frac{\sqrt{m}v_y}
{\sqrt{\hbar\Omega}}.
\end{equation}
In contrast, although both intra-$s$ and inter-$s$ channels remain active, the collapse of the lowest intra-$s$ excitation threshold makes this channel dominant at low energies. The resulting conductivity (Appendix~\ref{AppendixB}) reads
\begin{equation}\label{Eq10}
\sigma_{xx}(\Omega)=
\frac{e^{2}}{4\hbar}
\frac{2g_s}{5\pi G}
\frac{\sqrt{\hbar\Omega}}
{\sqrt{m}v_y}.
\end{equation}
This behavior results from a competition between the singular low-energy JDOS and the energy dependence of the optical matrix elements arising from the quadratic dispersion along the $k_x$ direction, where $v_x \sim k_x$, leading to $P_{\mathrm{eff},xx}^{-\rightarrow +}(\mathbf{k}) \propto \Omega$ as shown in Eq.~(\ref{EqB06}). As a result, the $\Omega^{-1/2}$ divergence is converted into the complementary scaling $\sigma_{xx}\propto \Omega^{1/2}$. These complementary power-law behaviors provide a characteristic optical signature of the semi-Dirac critical point and are consistent with previous theoretical studies of semi-Dirac systems~\cite{Carbotte2,Ziegler1,Mawrie1}.

At intermediate energies, deviations from the semi-Dirac critical scaling emerge due to the breakdown of the low-energy description beyond the vicinity of the band-touching region. In addition, the broadening and suppression of the inter-$s$ onset features in $\sigma_{xx}$ indicate a redistribution of optical spectral weight around the corresponding excitation thresholds, driven by the critical band reconstruction. At higher energies, the influence of the semi-Dirac critical spectrum gradually diminishes, allowing the system to recover its anisotropic linear bands and the corresponding constant optical conductivity values.

\begin{figure*}
\centering
\includegraphics[width=15cm,angle=0]{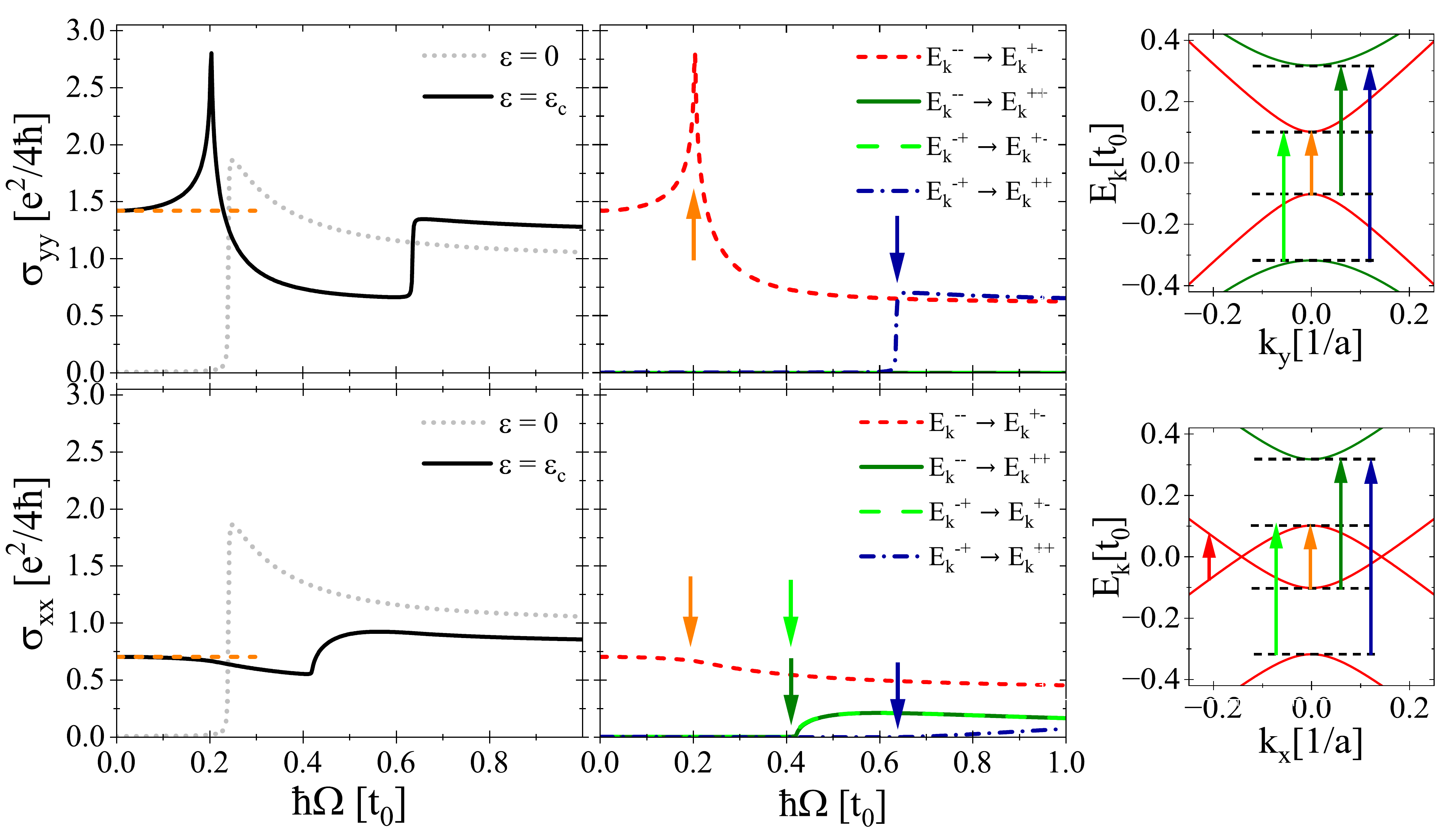}
\caption{\label{Fig06} Optical conductivity of strained Kekulé-O graphene in the post-critical regime ($\varepsilon=0.08$), where the semi-Dirac point splits into two anisotropic Dirac cones and saddle points emerge near the $\Gamma$ point. Left panels show $\sigma_{yy}$ (top) and $\sigma_{xx}$ (bottom), with gray dashed curves indicating the unstrained reference. Middle panels decompose the response into interband transition contributions, while right panels show the reconstructed low-energy band structure along $k_y$ and $k_x$ with corresponding transition pathways. The optical response combines a Dirac-like low-energy background with a strong van Hove resonance, whose pronounced polarization dependence originates from anisotropic velocity matrix elements rather than the joint density of states alone.}
\end{figure*}

Figure~\ref{Fig06} presents the longitudinal optical conductivity in the post-critical regime, characterized by the splitting of the semi-Dirac node into two anisotropic Dirac cones and the concomitant formation of saddle points near the $\Gamma$ point. The middle and right panels show the corresponding decomposition into intra-$s$ and inter-$s$ transition channels and the reconstructed band structure, respectively, following the notation introduced in Fig.~\ref{Fig04}.

In this regime, the optical response exhibits a crossover from a semi-Dirac critical scaling to a distinct behavior characterized by an anisotropic Dirac-like background at low energies and a pronounced van Hove resonance at finite energies. At low frequencies, both conductivity components are dominated by transitions between the gapless anisotropic Dirac bands, $E_{\mathbf{k}}^{--}\rightarrow E_{\mathbf{k}}^{-+}$, leading to an anisotropic constant optical conductivity, as reported in similar systems~\cite{Mohammadi5}. This low-energy behavior is well described by an effective theory (see Appendix~\ref{AppendixB}), which yields an analytical expression for the longitudinal optical conductivity,
\begin{equation}\label{Eq11}
\sigma_{\alpha\alpha}(\Omega)=
\frac{v_{\alpha}}{v_{\bar{\alpha}}} \frac{e^2}{4\hbar},
\end{equation}
where $\bar{\alpha}$ denotes the direction perpendicular to $\alpha$ ($x\leftrightarrow y$). For $\varepsilon=0.08$, the analytical results are in excellent agreement with the numerical data shown in Fig.~\ref{Fig06}.

The pronounced van Hove resonance emerges in $\sigma_{yy}$ due to the formation of saddle points in the energy spectrum. The channel-resolved decomposition (the middle panels of Fig.~\ref{Fig06}) shows that it originates from the transition $E_{\mathbf{k}}^{--}\rightarrow E_{\mathbf{k}}^{+-}$ involving states near the saddle-point region around $\Gamma$. The associated van Hove singularity enhances the JDOS and produces a strong resonance in $\sigma_{yy}$. In contrast, the same transition produces a dip-like feature in $\sigma_{xx}$. This indicates that while the saddle-point singularity enhances the joint density of states for both polarizations, it does not by itself determine the anisotropic response. The qualitative difference between $\sigma_{yy}$ and $\sigma_{xx}$ is therefore governed by the squared velocity matrix elements, which strongly enhance the $y$-polarized response while remaining comparatively suppressed for $x$-polarized light near the saddle points. At intermediate energies, a substantial redistribution of optical spectral weight suppresses and flattens the higher-energy absorption peaks for both $\sigma_{xx}$ and $\sigma_{yy}$. At higher energies, this effect fades, and the system recovers anisotropic linear bands, causing both conductivity components to saturate into distinct asymptotic values.

\begin{figure*}
\centering
\includegraphics[width=16cm,angle=0]{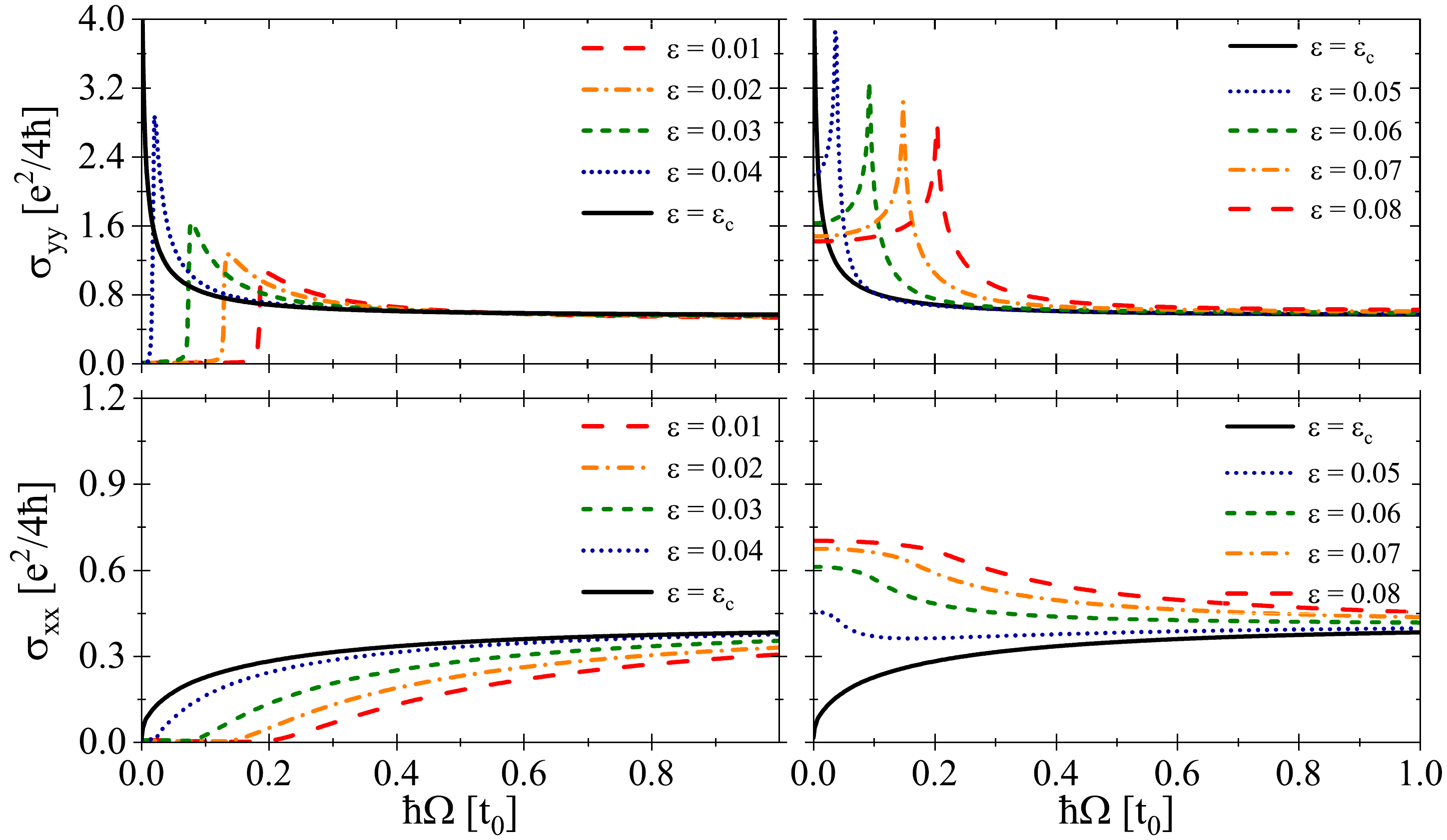}
\caption{\label{Fig07} Evolution of the first intra-$s$ interband contribution to the longitudinal optical conductivity of strained Kekulé-O graphene across the semi-Dirac transition. Upper panels show $\sigma_{yy}$ and lower panels show $\sigma_{xx}$. The left column corresponds to the gapped regime ($\varepsilon \le \varepsilon_c$), while the right column shows the post-critical regime ($\varepsilon > \varepsilon_c$). The optical response evolves from a gapped threshold behavior to semi-Dirac scaling at criticality, and then develops finite-energy structures associated with saddle-point transitions and van Hove singularities after Dirac-point splitting.}
\end{figure*}

Figure~\ref{Fig07} reveals how the characteristic low-energy optical fingerprints identified in Figs.~\ref{Fig04}--\ref{Fig06} emerge from the continuous evolution of the first intra-$s$ optical transition channel under strain. As strain increases, this channel evolves from a gapped absorption peak, through the semi-Dirac critical point, to a van Hove resonance associated with saddle-point formation following Dirac-point splitting. Consequently, the gapped absorption peak, the semi-Dirac critical response, and the van Hove optical resonance appear as successive manifestations of the same underlying optical transition channel reshaped by strain-induced band reconstruction.

For $\sigma_{yy}$, the evolution of the first intra-$s$ channel directly reflects the transformation of the underlying electronic structure across the strain-induced band reconstruction. In the gapped regime, the optical response exhibits a low-energy threshold peak associated with the gapped absorption edge. As strain increases and the gap decreases, this peak shifts toward lower frequencies while gaining spectral weight due to the increasing low-energy phase space available for optical excitations. At the critical strain, the absorption threshold collapses to zero energy and the response continuously evolves into the characteristic semi-Dirac scaling $\sigma_{yy}(\Omega)\propto\Omega^{-1/2}$, which, as discussed in Fig.~\ref{Fig05}, originates from the enhanced JDOS associated with the coexistence of linear and quadratic dispersions at the semi-Dirac critical point.

Beyond the critical strain, the singular low-energy response is transferred to finite energy as the semi-Dirac node splits into two anisotropic Dirac cones. Simultaneously, the band extrema near the $\Gamma$ point evolve into saddle points, producing a van Hove singularity in the electronic spectrum. The resulting enhancement of the JDOS generates the pronounced van Hove optical resonance identified in Fig.~\ref{Fig06}. Thus, the gapped threshold peak, the semi-Dirac critical response, and the post-critical van Hove resonance represent successive manifestations of the same underlying optical transition channel.

Although the same intra-$s$ channel is involved, its manifestation in $\sigma_{xx}$ is qualitatively different owing to the strong anisotropy of the velocity matrix elements. In the gapped regime, this intra-$s$ channel contributes only a relatively weak threshold structure. At the critical strain, this response evolves into the characteristic semi-Dirac behavior $\sigma_{xx}(\Omega)\propto\Omega^{1/2}$, reflecting the interplay between the critical phase space and the momentum dependence of the squared velocity matrix elements near the band-touching point.

After Dirac-point splitting, the semi-Dirac response is converted into the broad dip-like feature identified in Fig.~\ref{Fig06}. This dip occurs at the same characteristic energy scale as the van Hove resonance in $\sigma_{yy}$, indicating a common origin in saddle-point interband transitions. The differing spectral shapes in $\sigma_{xx}$ and $\sigma_{yy}$ arise from anisotropic velocity matrix elements, which enhance the $y$-polarized response near saddle points while suppressing the corresponding $x$-polarized contribution. As a result, the resonance and dip represent distinct manifestations of the same underlying transition processes in different polarizations. A nearly constant low-energy conductivity is also present in both components below the van Hove scale, originating from intraband processes of the reconstructed Dirac cones and corresponding to the Dirac-like background identified in the post-critical regime.

\begin{figure*}
\centering
\includegraphics[width=16cm,angle=0]{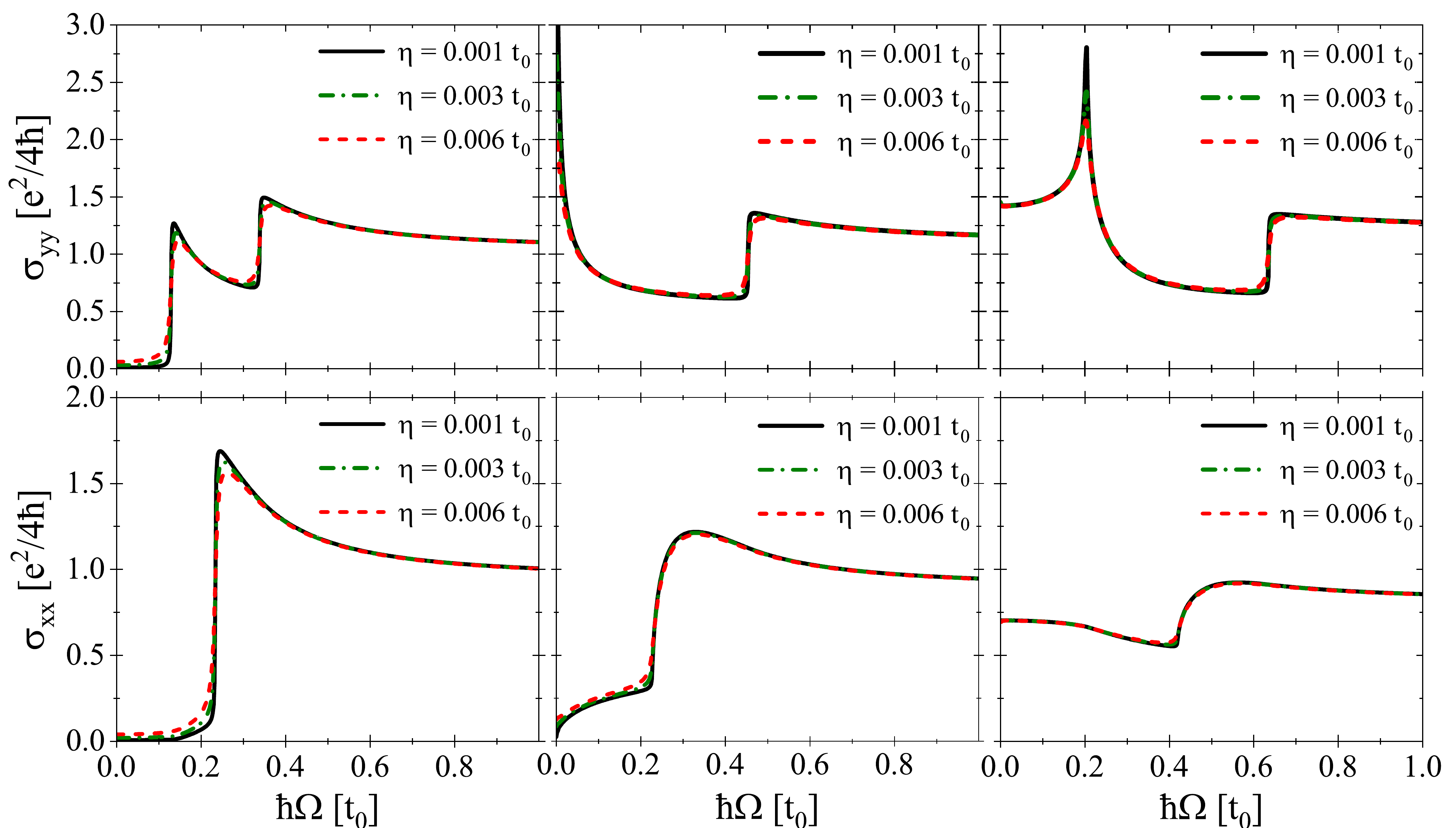}
\caption{\label{Fig08} Effect of broadening on the longitudinal optical conductivity of strained Kekulé-O graphene in the gapped, semi-Dirac critical-point, and post-critical regimes. Upper and lower panels correspond to $\sigma_{yy}$ and $\sigma_{xx}$, respectively. The conductivity is shown for three values of the phenomenological broadening parameter $\eta = 0.001t_0, 0.003t_0$, and $0.006t_0$ at $T=0$. The main anisotropic optical fingerprints remain well defined, with progressive smoothing of sharp interband structures as $\eta$ increases.}
\end{figure*}

Figure~\ref{Fig08} examines the stability of the optical fingerprints identified in Figs.~\ref{Fig04}--\ref{Fig06} against broadening effects. The effect of disorder is incorporated phenomenologically through the broadening parameter $\eta$, which represents a finite quasiparticle lifetime. The conductivity is shown for $\eta=0.001t_0$, $0.003t_0$, and $0.006t_0$ at zero temperature across the gapped regime, semi-Dirac critical point, and post-critical regime.

In the gapped regime, the finite-energy absorption thresholds are gradually broadened with increasing $\eta$. As a consequence, a small but finite optical conductivity develops below the nominal absorption edge, consistent with the Lorentzian representation of the Dirac delta functions used in the numerical calculations. Meanwhile, the sharp threshold peaks become rounded. Despite this broadening, the characteristic threshold energies remain essentially unaffected within numerical accuracy, indicating that the underlying interband excitation energies are only weakly affected by the finite quasiparticle lifetime associated with disorder.

The most pronounced effects of disorder-induced broadening occur near the semi-Dirac critical point, where the ideal low-energy power-law singularities are progressively rounded and regularized. For $\sigma_{xx}$, the critical behavior $\sigma_{xx}\propto\Omega^{1/2}$ is modified at the lowest frequencies, leading to a finite conductivity as $\Omega\rightarrow0$. For $\sigma_{yy}$, the divergent response $\sigma_{yy}\propto\Omega^{-1/2}$ is similarly suppressed, and the low-energy singularity is regularized into a finite broadened peak structure. Nevertheless, the characteristic anisotropic critical response remains well defined, demonstrating that the optical signature of the semi-Dirac point is preserved in the presence of moderate broadening.

In the post-critical regime, the low-energy Dirac-like background is weakly affected by increasing $\eta$. The van Hove optical resonance in $\sigma_{yy}$ and the corresponding dip-like structure in $\sigma_{xx}$ are only moderately broadened, while their characteristic energy scales remain essentially unchanged. This behavior reflects their origin in the strain-induced band reconstruction and the associated saddle-point structure of the electronic spectrum.

These results indicate that disorder-induced broadening primarily smooths the sharpest low-energy spectral singularities while preserving the characteristic energy scales and optical fingerprints of the three strain regimes. The gapped absorption peaks, the semi-Dirac features, and the saddle-point-induced van Hove resonance all remain well defined up to $\eta=0.006t_0$.

\begin{figure*}
\centering
\includegraphics[width=16cm,angle=0]{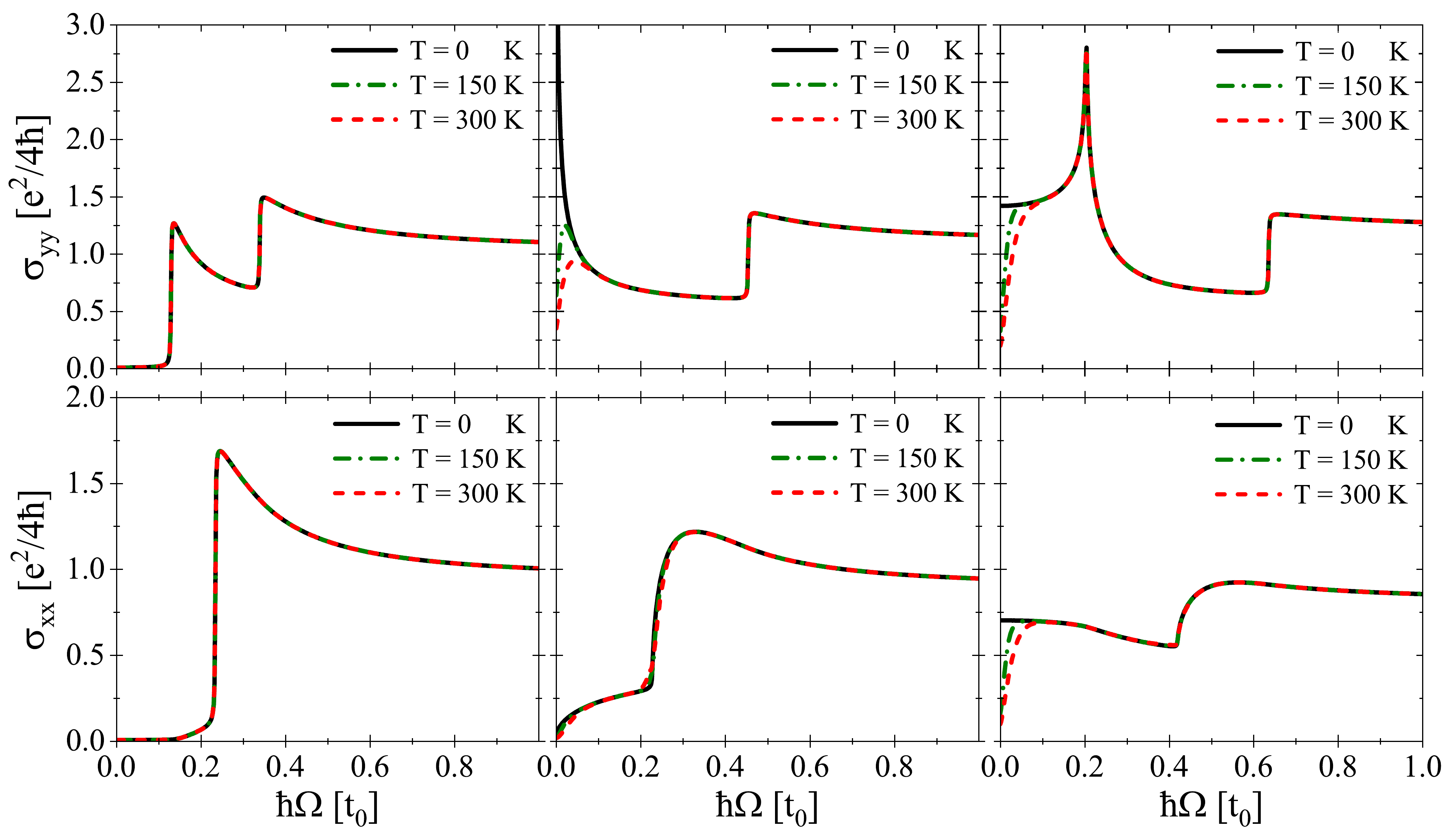}
\caption{\label{Fig09} Temperature dependence of the longitudinal optical conductivity of strained Kekulé-O graphene in the gapped, semi-Dirac critical-point, and post-critical regimes. Upper panels show $\sigma_{yy}$ and lower panels show $\sigma_{xx}$. Results are shown for $T=0$, $150$, and $300~\mathrm{K}$ with fixed impurity broadening $\eta=0.001t_0$. Thermal effects primarily smooth low-energy spectral features, while the characteristic optical fingerprints and polarization anisotropy remain clearly identifiable across all regimes.}
\end{figure*}

Figure~\ref{Fig09} examines the temperature dependence of the optical fingerprints identified in Figs.~\ref{Fig04}--\ref{Fig06}. The conductivity is shown for $T=0$, $150$, and $300~\mathrm{K}$ with fixed broadening $\eta=0.001t_0$ across the gapped regime, semi-Dirac critical point, and post-critical regime. Temperature enters the calculation exclusively through the Fermi--Dirac distribution function, while the electronic band structure and velocity matrix elements are assumed to be temperature independent. Having established the robustness of the optical fingerprints against disorder-induced broadening in Fig.~\ref{Fig08}, we now examine their response to thermal effects.

In the gapped regime, the optical response exhibits a well-defined interband absorption threshold associated with the band gap. The onset of optical conductivity is only weakly affected by temperature, indicating that the characteristic transition energies remain substantially larger than the thermal energy scale $k_B T$ throughout the temperature range considered. Consequently, thermal occupation effects produce only minor modifications of the low-energy spectral weight, while the threshold fingerprint remains clearly preserved. This weak temperature sensitivity is expected to diminish as the strain approaches regimes with smaller gap magnitudes, where the characteristic energy scale becomes comparable to $k_B T$.

The thermal response is most pronounced at the semi-Dirac critical point. Here, the low-energy optical conductivity is governed by the critical power-law behavior discussed in Appendix~\ref{AppendixB}. As temperature increases, the singular low-energy response undergoes significant thermal rounding. The effect is particularly strong in $\sigma_{yy}$, where the critical divergence $\sigma_{yy}\propto\Omega^{-1/2}$ is progressively suppressed and evolves into a finite broadened peak. In contrast, the corresponding $\sigma_{xx}\propto\Omega^{1/2}$ behavior remains comparatively insensitive to temperature. At low energies, the temperature dependence of the conductivity is well described by the analytical expressions in Eq.~(\ref{EqB09}), where thermal effects enter through the Fermi occupation factor and govern the low-energy response. This marked anisotropy originates from the distinct energy scaling of the two critical conductivity components, while both share the same thermal occupation factor, and highlights the strong sensitivity of the semi-Dirac fingerprint to thermal broadening effects.

In the post-critical regime, the constant Dirac-like background at low energies is significantly suppressed with increasing temperature. This behavior is driven by the thermal redistribution of the Fermi occupation factors, as analytically captured by Eq.~(\ref{EqB13}) and consistent with conventional graphene~\cite{Peres3,Stauber1}. In contrast, the van Hove optical resonance in $\sigma_{yy}$ remains remarkably robust against thermal effects, exhibiting only a minor reduction in intensity and slight broadening at $T = 300$ K while preserving its characteristic energy scale. Similar robustness is observed for the high-energy optical fingerprints.

Taken together, the temperature dependence reveals markedly different thermal robustness among the optical fingerprints. While the singular semi-Dirac critical response in $\sigma_{yy}$ is significantly rounded by thermal occupation effects, the gapped absorption threshold and the van Hove optical fingerprint remain robust against thermal effects up to room temperature. The characteristic polarization anisotropy and the principal optical fingerprints of the strain-driven electronic regimes therefore remain experimentally observable under realistic conditions.


Finally, recent experimental advances suggest that the strain-induced optical phenomena predicted here should be experimentally accessible. Kekulé-O bond ordering has already been realized in lithium-intercalated graphene~\cite{Bao1}, providing a realistic platform for investigating the electronic reconstruction associated with Kekulé ordering.

Controlled strain can be introduced using a variety of established techniques, including stressed metallic pads, thin-film overlayers, and local probe methods such as AFM nano-indentation~\cite{Shioya1,Shioya2,Mohiuddin1,Lee1}. These approaches have already been successfully employed to generate substantial and controllable strain profiles in graphene-based systems. Such methodologies offer experimentally feasible routes for tuning the system across the gapped regime, semi-Dirac critical point, and post-critical regime identified in this work.

The predicted optical fingerprints—including the anisotropic low-energy conductivity, the semi-Dirac critical response, and the van Hove optical resonance—can be probed directly through polarization-resolved optical spectroscopy~\cite{Naumis1,ahn1,mennel1}. Together with their demonstrated robustness against moderate broadening and finite temperatures, these results suggest that the principal optical fingerprints of the strain-driven electronic reconstruction in Kekulé-O graphene should remain experimentally observable under realistic conditions.

\section{\label{sec:Conclusions} Conclusions}

In this work, we have shown that strain provides an effective route for driving Kekulé-O graphene through a semi-Dirac transition, accompanied by a qualitative reorganization of its low-energy optical response. As the electronic spectrum evolves under strain, optical spectral weight is progressively transferred among the dominant interband excitation channels, resulting in a substantial enhancement of the polarization dependence of the conductivity. Guided by comprehensive numerical calculations based on the full four-band model and corroborated by analytical low-energy results, we have identified three distinct strain-induced regimes characterized by a gapped absorption threshold, complementary semi-Dirac critical scaling, and a post-critical regime exhibiting a constant low-energy conductivity background followed by a pronounced van Hove optical resonance.

In the gapped regime, the optical response is characterized by a pronounced polarization-dependent absorption threshold arising from the strain-induced anisotropy of the electronic spectrum. Consequently, $\sigma_{xx}$ and $\sigma_{yy}$ exhibit markedly different low-energy spectral profiles and threshold behaviors, providing the first optical fingerprint of the strain-driven electronic reconstruction.

At the critical strain, where the gap closes at the $\Gamma$ point, the system realizes a semi-Dirac spectrum with quadratic dispersion along one momentum direction and linear dispersion along the orthogonal direction. This anisotropic criticality gives rise to unconventional optical scaling laws, where $\sigma_{xx}(\Omega)\propto \Omega^{1/2}$ and $\sigma_{yy}(\Omega)\propto \Omega^{-1/2}$. The analytical expressions derived from the effective low-energy Hamiltonian are in excellent agreement with the full numerical results at low energies, establishing a direct connection between the anisotropic phase space, velocity matrix elements, and the emergent optical response.

Beyond the critical strain, the semi-Dirac point splits into two anisotropic Dirac cones, giving rise to a constant low-energy conductivity background associated with the reconstructed Dirac spectrum. In addition, saddle points emerge near the $\Gamma$ point, producing a pronounced optical resonance associated with a van Hove singularity. Together, the anisotropic Dirac-like background and the van Hove optical resonance constitute the characteristic low-energy optical fingerprint of the post-critical regime.

A central result of this work is that the low-energy optical fingerprints across the strain-driven transition can be understood within a unified framework based on the continuous evolution of a dominant interband transition channel. As strain reconstructs the electronic structure, this channel evolves from a gapped absorption edge to the semi-Dirac critical response and ultimately to the van Hove resonance, accompanied by a continuous redistribution of optical spectral weight.

We have further shown that these low-energy optical fingerprints exhibit varying degrees of robustness against impurity broadening and finite-temperature effects. While the singular semi-Dirac critical response is highly sensitive to thermal rounding, the gapped absorption threshold, the van Hove resonance, and the high-energy fingerprints remain robust under experimentally realistic conditions.

Together with recent experimental realizations of Kekulé ordering in graphene and established strain-engineering techniques, these findings indicate that the principal low-energy optical fingerprints remain experimentally accessible. Strained Kekulé-O graphene therefore provides a promising platform for exploring strain-controlled optical phenomena and anisotropic critical electronic behavior in two-dimensional materials.

\begin{acknowledgments}
This work was supported by Farhangian University.
\end{acknowledgments}

\renewcommand{\theequation}{A\arabic{equation}}
\setcounter{equation}{0}

\renewcommand{\thefigure}{A\arabic{figure}}
\setcounter{figure}{0}

\appendix

\section{\label{AppendixA} Low-energy effective model of strained Kekulé-O graphene}

In this Appendix, we derive the low-energy effective Hamiltonian of strained Kekulé-O graphene starting from the full $4\times4$ continuum model of Ref.~\cite{Andrade3}. We analyze the spectrum under uniaxial strain applied along the zigzag ($x$) direction, focusing on the principal momentum directions $k_x$ and $k_y$.

For strains smaller than or equal to the critical value ($\varepsilon \le \varepsilon_c$), the low-energy excitations are centered around the $\Gamma$ point and are fully captured by an expansion around $\mathbf{k}=0$. In contrast, for $\varepsilon > \varepsilon_c$, the low-energy physics is governed by newly emerged Dirac points located at $\pm K_{D}$ (with $K_{D}$ denoting the strain-induced Dirac points). The low-energy dispersion is obtained by restricting to the lower band ($s=-$) of the full four-band model and can be analyzed along the principal directions.

We first focus on the regime $\varepsilon \le \varepsilon_c$. In this case, the dispersion, for $k_x=0$, takes the form
\begin{equation}\label{EqA01}
E^{\lambda,-}_{k_{y},k_{x}=0} = \lambda \sqrt{(\hbar v_{y}k_{y})^{2} + \mathrm{E}_{g}^{2}},
\end{equation}
where
\begin{equation}\label{EqA02}
v_{y}=[1-\nu\varepsilon(1-\beta)- a A_{x}\Delta]v_{F},
\end{equation}
and
\begin{equation}\label{EqA03}
\mathrm{E}_{g}=3\tilde{t}_{0}\Delta-\hbar v_{F}A_{x}.
\end{equation}
Here, $A_x$ denotes the strain-induced gauge field associated with uniaxial deformation along the zigzag ($x$) direction. The parameter $\mathrm{E}_{g}$ acts as an effective mass term controlling the spectral gap, whose vanishing at $\varepsilon=\varepsilon_c$ signals the closing of the gap at the $\Gamma$ point. The dispersion along $k_y$ thus retains a Dirac-like character with a strain-dependent gap.

While the dispersion along $k_y$ remains Dirac-like, the behavior along $k_x$ reveals a fundamentally different, nonrelativistic character. For $k_y=0$, the dispersion is given by
\begin{equation}\label{EqA04}
E^{\lambda,-}_{k_{x},k_{y}=0} = \lambda \Big( \sqrt{ (\hbar v_{F}\bar{k}_{x})^{2} + (3\tilde{t}_{0}\Delta)^{2} } - \hbar v_{F}A_{x} \Big),
\end{equation}
where $\bar{k}_x$ is the strain--modified momentum along the $x$ direction. For $\varepsilon \le \varepsilon_c$ and in the low-energy regime around the $\Gamma$ point, defined by $\hbar v_{F}\bar{k}_{x} \ll 3\tilde{t}_{0}\Delta$, expanding the square root yields
\begin{equation}\label{EqA05}
\sqrt{ (\hbar v_{F}\bar{k}_{x})^{2} + (3\tilde{t}_{0}\Delta)^{2} }
\simeq
3\tilde{t}_{0}\Delta + \frac{\hbar^{2} v_{F}^{2} \bar{k}_{x}^{2}}{6\tilde{t}_{0}\Delta}.
\end{equation}
Substituting into the dispersion, one obtains
\begin{equation}\label{EqA06}
E^{\lambda,-}_{k_{x},k_{y}=0} = \lambda \Big( \frac{\hbar^{2} k_{x}^{2}}{2m} + \mathrm{E}_{g} \Big),
\end{equation}
with
\begin{equation}\label{EqA07}
\frac{1}{2m}=\frac{([1+\varepsilon(1-\beta)]v_{F})^{2}}{6\tilde{t}_{0}\Delta}.
\end{equation}
Thus, for $\varepsilon \le \varepsilon_c$, the low-energy spectrum near the $\Gamma$ point is strongly anisotropic: it is linear along $k_y$, while it exhibits a gapped, approximately quadratic behavior along $k_x$, which becomes purely quadratic at $\varepsilon=\varepsilon_c$ where $\mathrm{E}_{g}=0$, corresponding to a semi-Dirac dispersion at the critical strain.
The anisotropic dispersions motivate the following effective Hamiltonian near the $\Gamma$ point:
\begin{equation}\label{EqA08}
\hat{H}_{\mathrm{eff}} = \left( \frac{\hbar^2 k_x^2}{2 m} + \mathrm{E}_{g} \right)\sigma_x + \hbar v_y k_y \sigma_y,
\end{equation}
which yields the eigenvalues
\begin{equation}\label{EqA09}
E_{\mathbf{k}}^{\lambda} = \lambda \sqrt{ (\hbar^2 k_x^2 / 2m + \mathrm{E}_{g})^2 + (\hbar v_y k_y)^2 }.
\end{equation}
Comparison our numerical recasts with earlier studies~\cite{Carbotte2} further demonstrates that the optical conductivity along the $x$ direction lies beyond the scope of a simple gapped-Dirac description. Consequently, the optical response cannot be captured by an effective massive Dirac Hamiltonian.

We now consider the regime $\varepsilon > \varepsilon_c$, where the low-energy spectrum is governed by strain-induced Dirac points at $\pm K_{D}$ along the $k_x$ direction, where
\begin{equation}\label{EqA10}
K_{D}=\frac{\sqrt{|\mathrm{E}_{g}|(\hbar v_{F}A_{x}+3\tilde{t}_{0}\Delta)}}{\hbar[1+(1-\beta)\varepsilon]v_{F}}.
\end{equation}
We expand the spectrum around the strain-induced Dirac points $\pm \bm{K}_{D}$ for $|\bm{k}\mp\bm{K}_{D}| \ll |\bm{K}_{D}|$ and retain the leading terms. The resulting dispersion takes the form
\begin{equation}\label{EqA11}
E^{\lambda}_{\mathbf{k}}=\lambda \hbar\sqrt{v_{x}^{2}(k_{x}\mp K_{D})^{2} + v_{y}^{2}k_{y}^{2}},
\end{equation}
where
\begin{subequations}\label{EqA12}
\begin{align}
v_{x}=\frac{[1+\varepsilon(1-\beta)]^{2}v_{F}K_{D}}{A_{x}},\\
v_{y}=[1-\nu\varepsilon(1-\beta)]v_{F}-\frac{3a\tilde{t}_{0}\Delta^{2}}{\hbar}.
\end{align}
\end{subequations}
The above linear dispersion motivates the following effective Hamiltonian in the vicinity of $\pm K_{D}$:
\begin{equation}\label{EqA13}
\hat{H}_{\mathrm{eff}} =
\hbar v_x (k_x \mp K_{D})\, \sigma_x
+ \hbar v_y k_y \, \sigma_y .
\end{equation}
Here, the upper (lower) sign corresponds to the Dirac point at $+K_{D}$ ($-K_{D}$), and $\sigma_{x,y}$ are Pauli matrices acting in the sublattice pseudospin space.

While the preceding derivations consider uniaxial strain applied along the zigzag ($x$) direction, the low-energy effective description remains qualitatively similar for strain applied along the armchair ($y$) direction, albeit with renormalized parameters. For the Hamiltonian near the $\Gamma$ point [Eq.~(\ref{EqA08})], the mass and velocity parameters are modified as:
\begin{equation}\label{EqA14}
\frac{1}{2m} = \frac{\left([1-\nu\varepsilon(1-\beta)]v_F\right)^2}{6 \tilde{t}_0 \Delta},
\end{equation}
and
\begin{equation}\label{EqA15}
v_y = \left[1 + \varepsilon(1-\beta) - a A_x \Delta\right] v_F,
\end{equation}
whereas the effective mass term $\mathrm{E}_{g}$ remains invariant under this change of orientation. Similarly, for the split-Dirac regime
($\varepsilon > \varepsilon_c$), the Fermi velocities near $\pm K_D$ [Eq.~(\ref{EqA12})] are renormalized to:
\begin{subequations}\label{EqA16}
\begin{align}
v_{x}=\frac{[1-\nu\varepsilon(1-\beta)]^{2}v_{F}K_{D}}{A_{x}}, \\
v_{y}=[1+\varepsilon(1-\beta)]v_{F}-\frac{3a\tilde{t}_{0}\Delta^{2}}{\hbar}.
\end{align}
\end{subequations}
These adjustments account for the specific geometry of the pseudo-gauge fields under armchair deformation~\cite{Pereira1}, while preserving the overall semi-Dirac character of the transition.

\renewcommand{\theequation}{B\arabic{equation}}
\setcounter{equation}{0}

\renewcommand{\thefigure}{B\arabic{figure}}
\setcounter{figure}{0}


\section{\label{AppendixB}Analytical derivation of the optical conductivity within the low-energy model}

We evaluate the longitudinal optical conductivity within the Kubo formalism starting from the effective Hamiltonian derived in Appendix~\ref{AppendixA}. Depending on the strain regime, the system exhibits distinct anisotropic electronic phases. In this appendix, we provide a controlled derivation of the optical response using the corresponding low-energy descriptions.

\vspace{0.3cm}

\textbf{The semi-Dirac critical point.}
In this regime, the low-energy physics is centered at the $\Gamma$ point and is described by the effective Hamiltonian in Eq.~\ref{EqA08}. The spectral function is written in terms of band-projection matrices as
\begin{equation}\label{EqB01}
\mathcal{\hat{A}}_{\mathrm{eff}}(\omega,\bm{k}) = 2\pi \sum_{\lambda} \hat{M}_{\mathrm{eff}}^{\lambda}(\bm{k})\,
\delta\!\left(\hbar\omega - E_{\bm{k}}^{\lambda}\right).
\end{equation}
To evaluate the Kubo formula analytically, we introduce the anisotropic energy parametrization $\epsilon_x = \frac{\hbar^2 k_x^2}{2m}$, $\epsilon_y = \hbar v_y k_y$, and define
\begin{equation}\label{EqB02}
\epsilon = \sqrt{\epsilon_x^2 + \epsilon_y^2}, \qquad \tan\phi = \frac{\epsilon_y}{\epsilon_x}.
\end{equation}
This mapping is one-to-one in the sector $k_x \ge 0,\quad k_y \in (-\infty,\infty)$, while the contribution from $k_x < 0$ is included via a symmetry factor of two. Within this parametrization, the momentum measure becomes
\begin{equation}\label{EqB03}
dk_x dk_y =
\frac{\sqrt{2m}}{2\hbar^2 v_y}
\frac{\sqrt{\epsilon}}{\sqrt{\cos\phi}}\, d\epsilon d\phi.
\end{equation}
The apparent square-root singularity at $\phi \to \pm \pi/2$ is a coordinate artifact of the anisotropic transformation and does not correspond to any physical divergence in the density of states. The angular variable is restricted to $\phi \in [-\pi/2,\pi/2]$, and the full phase space is recovered by the symmetry factor mentioned above.
The band-projection matrices in this representation are
\begin{equation}\label{EqB04}
M_{11}^{\lambda} = M_{22}^{\lambda} = \frac{1}{2},
\end{equation}
and
\begin{equation}\label{EqB05}
M_{12}^{\lambda} = M_{21}^{\lambda \ast} =
\frac{ \cos\phi - i \sin\phi}{2}.
\end{equation}
The interband squared velocity matrix elements take the form
\begin{subequations}\label{EqB06}
\begin{align}
P_{\mathrm{eff},xx}^{-\rightarrow +}(\mathbf{k}) &= \frac{2 \epsilon \cos\phi \sin^2\phi}{m} , \\
P_{\mathrm{eff},yy}^{-\rightarrow +}(\mathbf{k}) &= v_{y}^{2} \cos^2\phi .
\end{align}
\end{subequations}
Using the identity
\begin{equation}\label{EqB07}
\int_{-\pi/2}^{\pi/2} d\phi \, \sin^m\phi \cos^n\phi
= \frac{\Gamma\left(\frac{m+1}{2}\right)\Gamma\left(\frac{n+1}{2}\right)}
{\Gamma\left(\frac{m+n+2}{2}\right)},
\end{equation}
the angular integrals reduce to a single geometric constant
\begin{equation}\label{EqB08}
G = \frac{1}{\pi} \int_{-\pi/2}^{\pi/2} \frac{d\phi}{\sqrt{\cos\phi}}
= \frac{1}{2\sqrt{\pi}} \frac{\Gamma(1/4)}{\Gamma(3/4)}.
\end{equation}
After performing the radial integration constrained by the delta function, the optical conductivity at the semi-Dirac critical point becomes
\begin{subequations}\label{EqB09}
\begin{align}
\sigma_{xx}(\Omega) &= \frac{e^{2}}{4\hbar}
\frac{2 g_s}{5\pi G}
\frac{\sqrt{\hbar\Omega}}{\sqrt{m} v_y}
\tanh\left(\frac{\hbar\Omega}{4k_B T}\right), \\
\sigma_{yy}(\Omega) &= \frac{e^{2}}{4\hbar}
\frac{g_s G}{3}
\frac{\sqrt{m} v_y}{\sqrt{\hbar\Omega}}
\tanh\left(\frac{\hbar\Omega}{4k_B T}\right),
\end{align}
\end{subequations}
providing a clear optical signature of the merging of Dirac points, the conductivity vanishes as $\sqrt{\Omega}$ in the $x$-direction while diverging as $1/\sqrt{\Omega}$ in the $y$-direction.

\vspace{0.3cm}

\textbf{The post-critical regime.}

For $\varepsilon > \varepsilon_c$, the spectrum reorganizes into two anisotropic Dirac cones located at $\pm K_D$. The interband transitions are governed by the band-projection matrix, $\hat{M}_{\mathrm{eff}}^{\lambda}(\bm{k})$, for the separated Dirac cones. After a shift of momentum coordinates, $q_{x} = k_{x} \mp K_{D}$ and $q_{y} = k_{y}$, the diagonal elements are $M_{11}^{\lambda}= M_{22}^{\lambda}=\frac{1}{2}$, while the off-diagonal terms are:
\begin{equation}\label{EqB10}
M_{12}^{\lambda}=M_{21}^{\lambda \ast}=\lambda\frac{ \pm\hbar v_{x} q_{x} - i \hbar v_{y} q_{y} }{2E_{\mathbf{q}}^{+}},
\end{equation}
where $E_{\mathbf{q}}^{+} = \sqrt{(\hbar v_x q_x)^2 + (\hbar v_y q_y)^2}$. The corresponding squared velocity matrix elements in this regime are:
\begin{equation}\label{EqB11}
P_{\mathrm{eff},\alpha\alpha}^{-\rightarrow +}(\bm{q}) = \left(\frac{\hbar v_{\alpha}v_{\bar{\alpha}} q_{\bar{\alpha}}}
{E_{\bm{q}}^{+}}\right)^{2},
\end{equation}
where $\bar{\alpha}$ is the direction perpendicular to $\alpha$ ($x \leftrightarrow y$). Substituting these into the Kubo formula [Eq.~(\ref{Eq09})], the optical conductivity at $T=0$ takes the form:
\begin{equation}\label{EqB12}
\begin{aligned}
\sigma_{\alpha\alpha}(\Omega) = &\frac{\pi g_s e^2}{\Omega} \int \frac{dq_{x}dq_{y}}{4\pi^2} \Big( \frac{\hbar v_{\alpha} v_{\bar{\alpha}}q_{\bar{\alpha}}}{E_{\mathbf{q}}^{+}} \Big)^{2} \\
&\tanh\left(\frac{E_{\mathbf{q}}^{+}}{2k_{B}T}\right) \delta\left(\hbar \Omega - 2E_{\mathbf{q}}^{+} \right).
\end{aligned}
\end{equation}
This yields the final anisotropic response:
\begin{equation}\label{EqB13}
\sigma_{\alpha\alpha}(\Omega) = \frac{v_{\alpha}}{v_{\bar{\alpha}}} \frac{e^2}{4\hbar}
\tanh\left(\frac{\hbar\Omega}{4k_{B}T}\right),
\end{equation}
This recovers the universal Dirac response with strain-induced anisotropy.

Note that the zero-temperature limits of the optical conductivity in Eqs.~(\ref{EqB09}) and~(\ref{EqB13}) are readily obtained by applying the limiting behavior of the hyperbolic tangent function at $T \to 0$ [i.e., $\lim_{T \to 0} \tanh(\frac{\hbar\Omega}{4k_B T}) = 1$ for $\Omega > 0$].

\nocite{*}

\bibliography{Refs}

\end{document}